\documentclass[preprint,12pt]{elsarticle}
\usepackage[english]{babel}
\usepackage{amssymb}
\usepackage{float}
\usepackage{amsmath}
\usepackage{subfigure}
\usepackage{epsfig}
\journal{NIM, Sect. A}
\usepackage{color} 

\begin{document}

\begin{frontmatter}

\title{Comparison of two analysis methods for nuclear reaction measurements of $^{12}$C +$^{12}$C interactions at 95~MeV/u for hadrontherapy}

\author[CAEN] {J. Dudouet}
\author[IPHC] {D. Juliani}  
\author[CAEN] {M. Labalme}
\author[CAEN] {J.C. Ang\'elique}
\author[CEA] {B. Braunn}
\author[CAEN] {J. Colin}
\author[CAEN] {D. Cussol}
\author[IPHC] {Ch. Finck} 
\author[CAEN] {J.M. Fontbonne}
\author[CAEN] {H. Gu\'erin}
\author[IPNL] {P. Henriquet}
\author[IPNL] {J. Krimmer}
\author[IPHC] {M. Rousseau} 
\author[GANIL] {M.G. Saint-Laurent}
\address[CAEN]{LPC Caen, ENSICAEN, Universit\'e de Caen, CNRS/IN2P3, Caen, France}
\address[IPHC]{Institut Pluridisciplinaire Hubert Curien Strasbourg, France}
\address[CEA]{CEA, Centre de Saclay, IRFU/SPhN, F-91191, Gif-sur-Yvette, France}
\address[IPNL]{IPNL, Universit\'e de Lyon, F-69003 Lyon, France; Universit\'e Lyon 1 and CNRS/IN2P3, UMR 5822 F-69622 Villeurbanne, France}
\address[GANIL]{Grand Acc\'el\'erateur National d'Ions Lourds (GANIL),CEA/DSM-CNRS/IN2P3 BP5027,F-14076 Caen cedex 5, France }
 
\begin{abstract}
\textit{During therapeutic treatment with heavier ions like carbon, the beam undergoes nuclear fragmentation and secondary light charged particles, in particular protons and alpha particles, are produced. To estimate the dose deposited into the tumors and the surrounding healthy tissues, the accuracy must be higher than ($\pm 3\%$ and $\pm$ 1~mm). Therefore, measurements  are performed to determine the double differential cross section for different reactions. In this paper, the analysis of data from  $^{12}$C +$^{12}$C reactions at 95~MeV/u are presented. The emitted particles are detected with $\Delta$E$_\text{thin}$-$\Delta$E$_\text{thick}$-E telescopes made of a stack of two silicon detectors and a CsI crystal. Two different methods are used to identify the particles. One is based on graphical cuts onto the $\Delta$E-E maps, the second is based on the so-called KaliVeda method using a functional description of $\Delta E$ versus $E$. The results of the two methods will be presented in this paper as well as the comparison between both.}

\end{abstract} 


\end{frontmatter}

\newpage

\section{Introduction}
Carbon ions with energies ranging from 80 to 400~MeV/u are used in hadrontherapy to treat cancerous tumors. Their advantage compared to photons lies in the great accuracy of ion beams to target the tumor while sparing the surrounding healthy tissues (due to the high dose deposition in the Bragg peak and the small angular scattering of ions).
The physical dose deposition is affected by the fragmentation of the ions along their penetration path in the patient tissues~\cite{Schardt 96}. This process leads to a consumption of the primary beam (up to 70\%  for 400~MeV/u $^{12}$C in water) and to a delocalisation of the dose in the healthy tissues~\cite{Scholz 00}.

Simulation codes are used to compute the transport of ions in matter, but the constraints on nuclear models and fragmentation cross sections in the energy range used in hadrontherapy (80 to 400~MeV/u) are not yet sufficient to reproduce the fragmentation processes with the required accuracy for clinical treatments. To improve the knowledge on the $^{12}$C fragmentation process, experiments have been performed in Japan and in Europe for more than 15 years. Measurements of light charged fragment production in water and PMMA (C$_5$H$_8$O$_2$) have been started by the Japanese treatment centers (Chiba and Hyogo) and by the Gesellschaft f\"ur SchwerIonenforschung (GSI) biophysics department in the 
energy range 200-400 MeV/u~\cite{Matsufuji 03}~\cite{Matsufuji 05}~\cite{Toshito 07}~\cite{Schall 96}~\cite{Gunzert 04}~\cite{Gunzert 08}. 

To extend these data to the lowest energies, a first integral experiment on thick water equivalent targets has been performed by our collaboration in May 2008 at the Grand Acc\'el\'erateur National d'Ions Lourds (GANIL, France). Energy and angular distributions of the fragments produced by nuclear reactions of 95 MeV/u $^{12}$C with thick PMMA targets have been obtained~\cite{Braunn 11}. Comparisons between experimental data and simulations, using different physics processes of GEANT4~\cite{geant} for modeling the passage of particles through matter, have shown discrepancies. Nevertheless, it is difficult to constrain the nuclear reaction models by a direct comparison to thick targets experiments.

To improve the models and reach the accuracy required for a reference simulation code for hadrontherapy, a second experiment has been performed on thin targets on May 2011 at GANIL to study C-C, C-H, C-O, C-Al and C-$^{nat}$Ti reactions at 95~MeV/u. The particle identification has been performed by using telescopes (stack of detectors) and the $\Delta$E-E technique. The experimental set-up will be presented in the next section.

In such experiments, the usual procedure to extract the particle natures and the particle energies from raw data consists in two separate steps: the particle identificaton and the energy calibration of the detectors. The particle identification is done by using graphical cuts on so called $\Delta$E-E maps. The energy calibration is done by using well defined reference energies (elastic energies, punch through energies). We propose in this paper an alternative procedure which performs the particle identification and the detector calibrations in one single step. The two different analysis methods will be described, the first one called "graphical method" uses graphical cuts, the second one called ``KaliVeda method'' uses the KaliVeda framework developed by the INDRA collaboration~\cite{Pouthas 95}\cite{kaliveda}. The main advantage of the KaliVeda method is that once the identification grids are created, they allow to do both the detector calibration and the particle identification at the same time thanks to automated procedures which saves time in data analysis. 
A comparison of the results obtained with the two methods for the C-C reaction will be presented in the last section. 

\section{Experimental setup}

\subsection{Charged particles detection}

\begin{figure}[H]
\subfigure[]{\label{eclan1}{\includegraphics[width=0.49\textwidth]{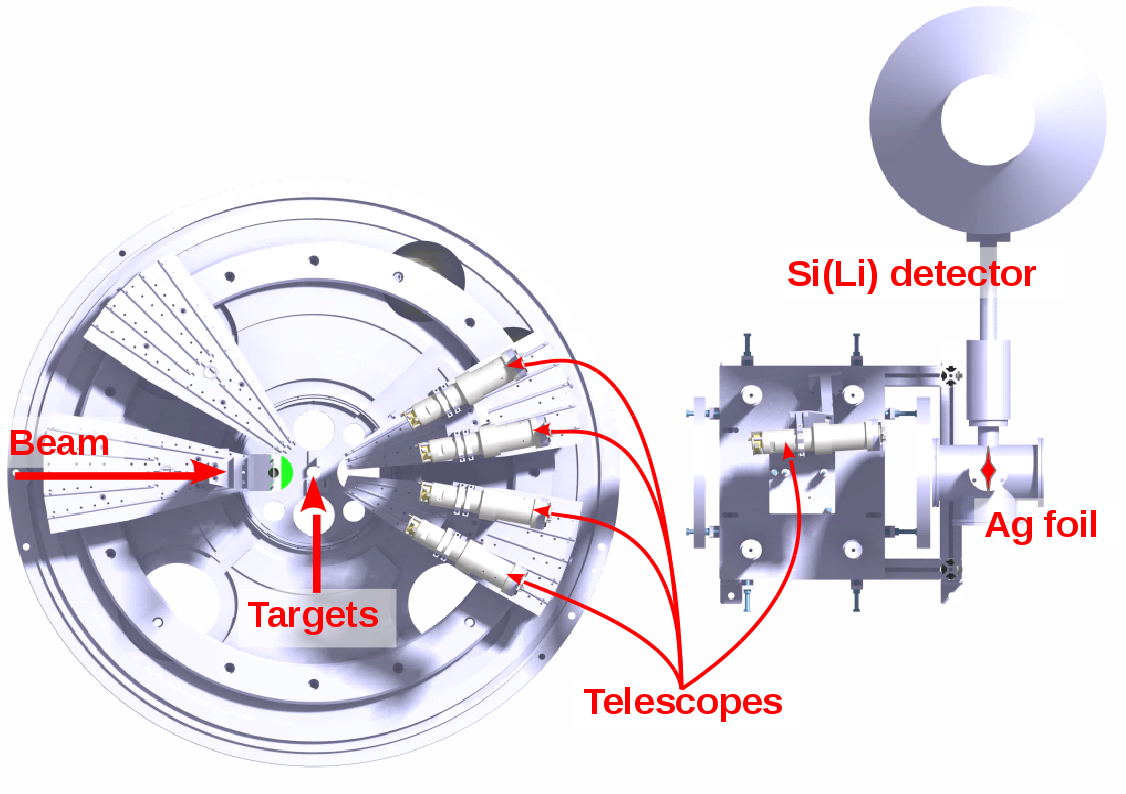}}}
\subfigure[]{\label{eclan2}{\includegraphics[width=0.49\textwidth]{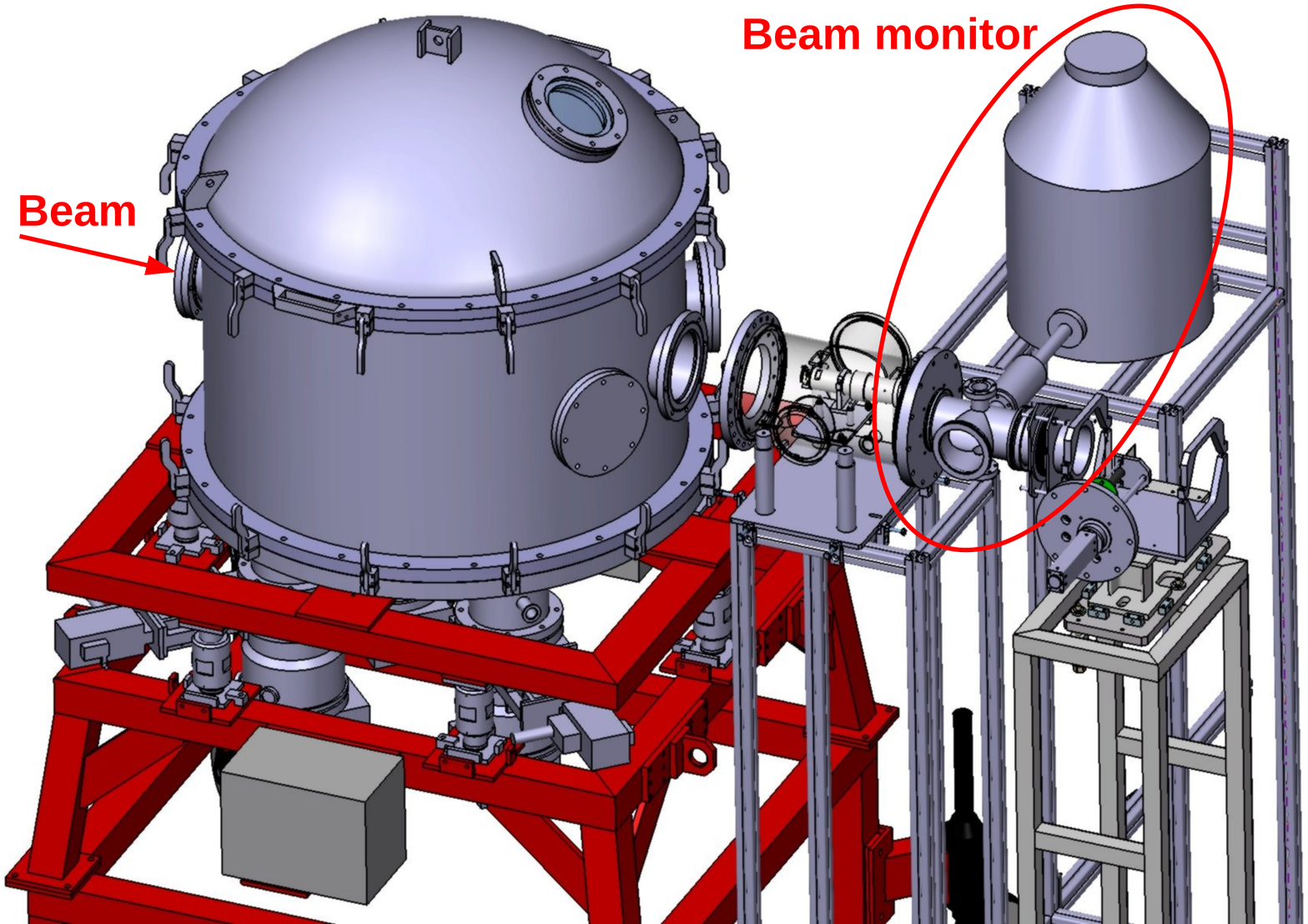}}}
\caption{Representation of the experimental setup.}
\label{eclan}
\end{figure}

The experiment has been performed using the ECLAN reaction chamber at the GANIL G22 beam line.
The energy of the impinging carbon ions was 95~MeV/u on different thin targets (C, CH$_2$, Al, Al$_2$O$_3$, $^{nat}$Ti)
with area densities of about $\sim$100~mg/cm$^2$. In this section we will report only about carbon target data.
The setup consists of four $\Delta$E$_{\text{thin}}$-$\Delta$E$_{\text{thick}}$-E telescopes mounted two by two on two rotating stages that allows rotation inside the chamber from $7^\circ$ to $43^\circ$ at a fixed distance of 22~cm behind the target. Thus the solid angle covered by each telescope was 6.5~msr. 
A fifth telescope was at fixed  angle  at $4^\circ$, downstream, covering a solid angle of 0.4~msr as shown in Fig.~\ref{eclan}.

The five $\Delta$E$_{\text{thin}}$-$\Delta$E$_{\text{thick}}$-E telescopes consist in two silicon detectors followed by a 10~cm thick CsI scintillator. 
The thicknesses of the silicon detectors were 150 $\mu$m and 1000~$\mu$m respectively.

 \subsection{Beam Monitor}
\begin{figure}[H]
\centering
\subfigure{\label{SiLi}{\includegraphics[width=0.49\textwidth]{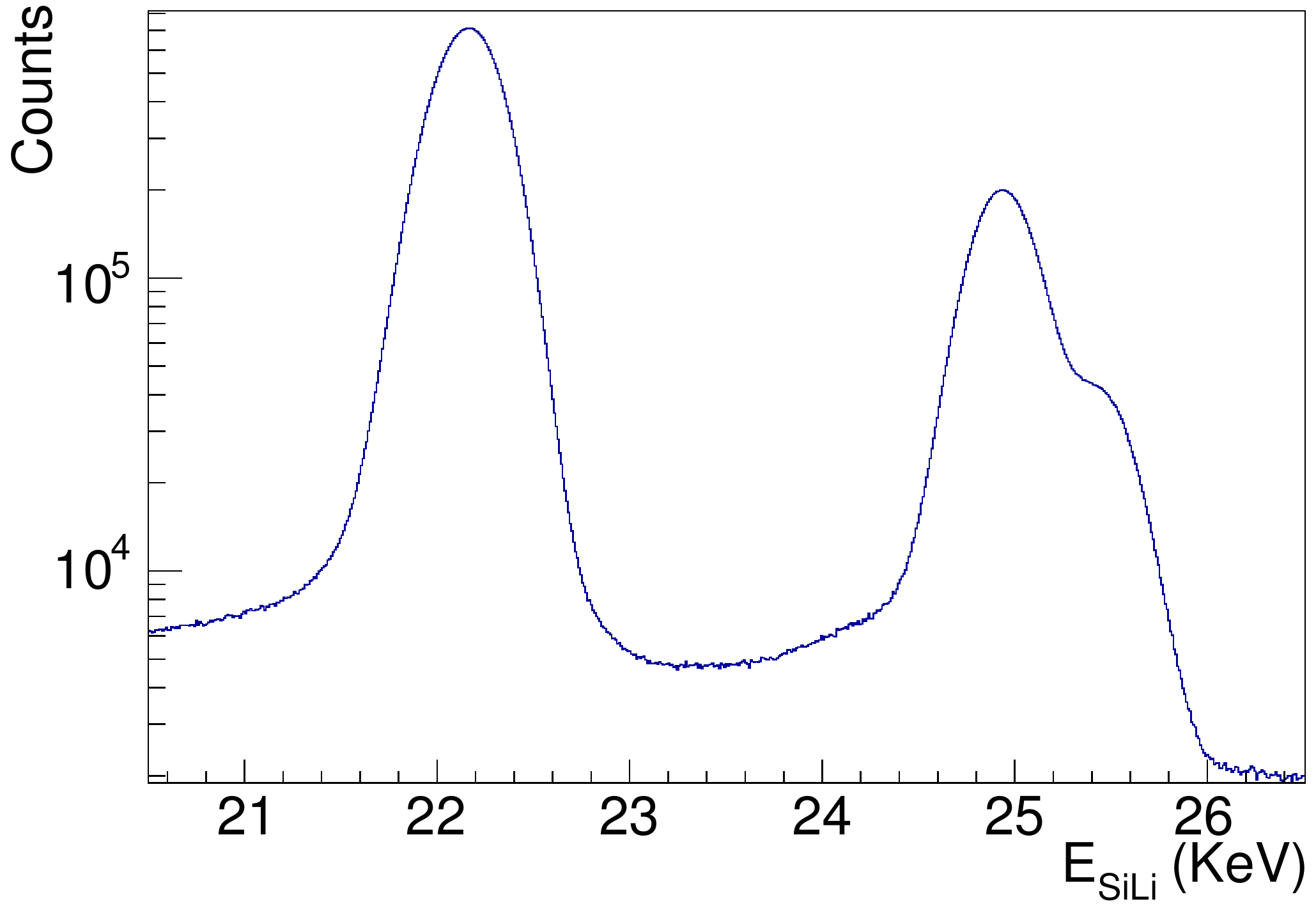}}}
\subfigure{\label{plast}{\includegraphics[width=0.49\textwidth]{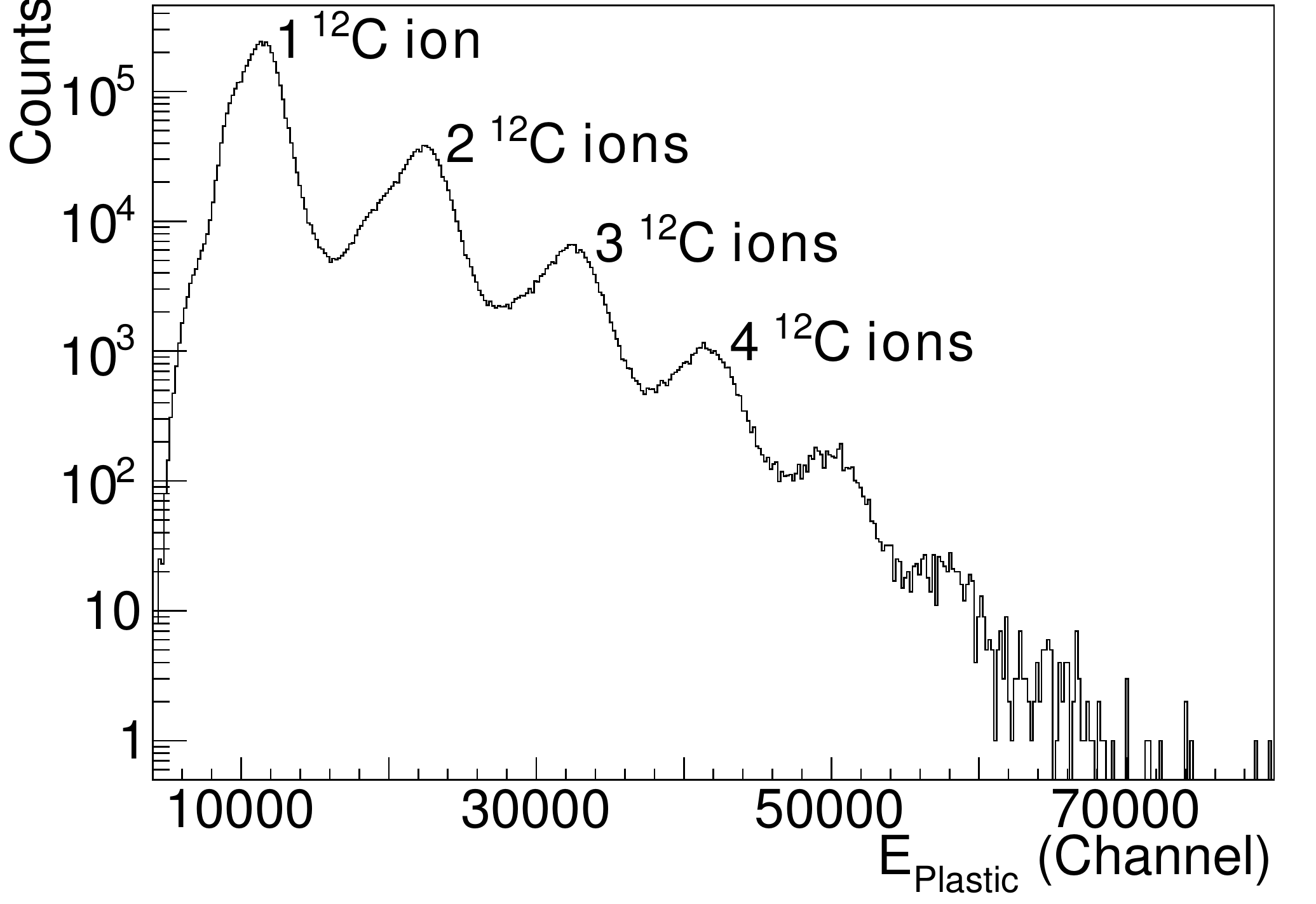}}}
\caption{Left-hand side: Energy spectrum of the X-rays emitted by the Ag foil obtained with a Si(Li) detector. Right-hand side: Energy spectrum obtained with the plastic scintillator for the beam monitor calibration.}
\end{figure}

\begin{figure}[H]
\centerline{\includegraphics[width=0.7\textwidth]{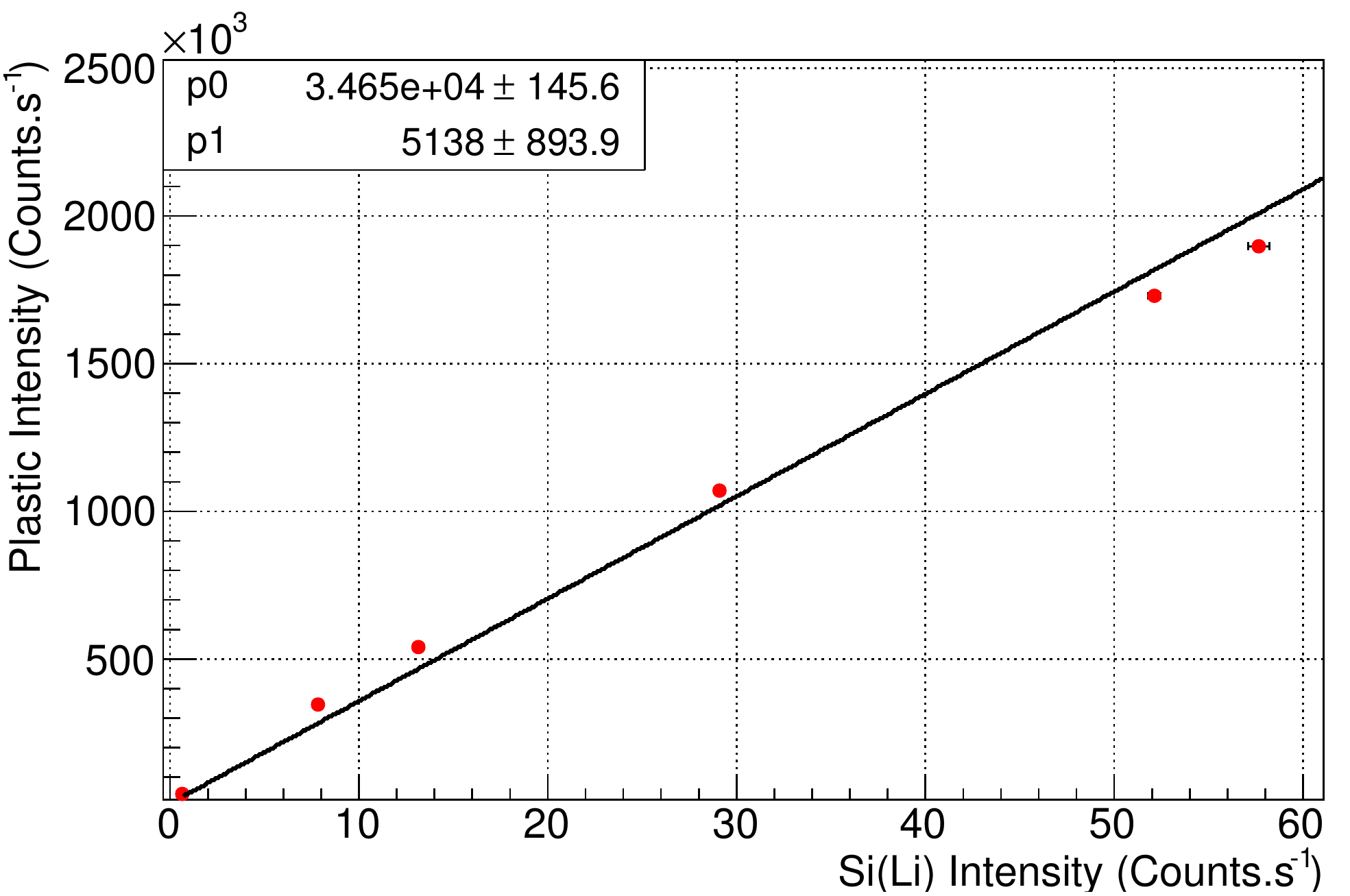}}
\caption{Intensity calibration of the SiLi detector thanks to the plastic scintillator intercepting the beam at low energy (I$_\text{plast}=$ p0$\times $I$_\text{Si(Li)}$ + p1).}
\label{etalplas}
\end{figure}

The beam monitoring is based on the measurement of fluorescence X-rays emitted by a thin Ag foil (7~$\mu$m thick) set in
the beam after the reaction chamber. X-rays are detected by means of two Si(Li) detectors located at $90^\circ$ with respect to the beam direction. These two detectors are calibrated at different beam intensities ranging from 10$^3$ to 10$^6$ ions/s by means of a plastic scintillator intercepting the beam. The beam intensity is determined with the two detectors (Si(Li)). The detected fluorescence X-rays, emitted by the thin Ag foil, are represented by two peaks in the energy spectra of the Si(Li) (Fig.~\ref{SiLi}). These peaks have been integrated while taking into account the background. Different background subtraction methods have been tested: no background, constant background, linear background, polynomial background. They are all compatible within 2\%. The energy spectrum of the plastic scintillator presents several peaks corresponding to the number of carbon ions detected in coincidence (Fig.~\ref{plast}). These peaks have been fitted with gaussian functions and integrated hereafter. Fig~\ref{etalplas} represents the beam intensity in the plastic detector relative to the detected X-rays intensity in one Si(Li) for several calibration runs. The calibration parameters are obtained by a linear fit of the calibration points.
This monitor gives a global accuracy of 5\% on the whole intensity range of the experiment (from 10$^5$ to 10$^7$ ions/s).

\section{Data Analysis using graphical cuts}

\subsection{Particle identification}
\begin{figure}[H]
\centerline{\includegraphics[width=0.7\textwidth]{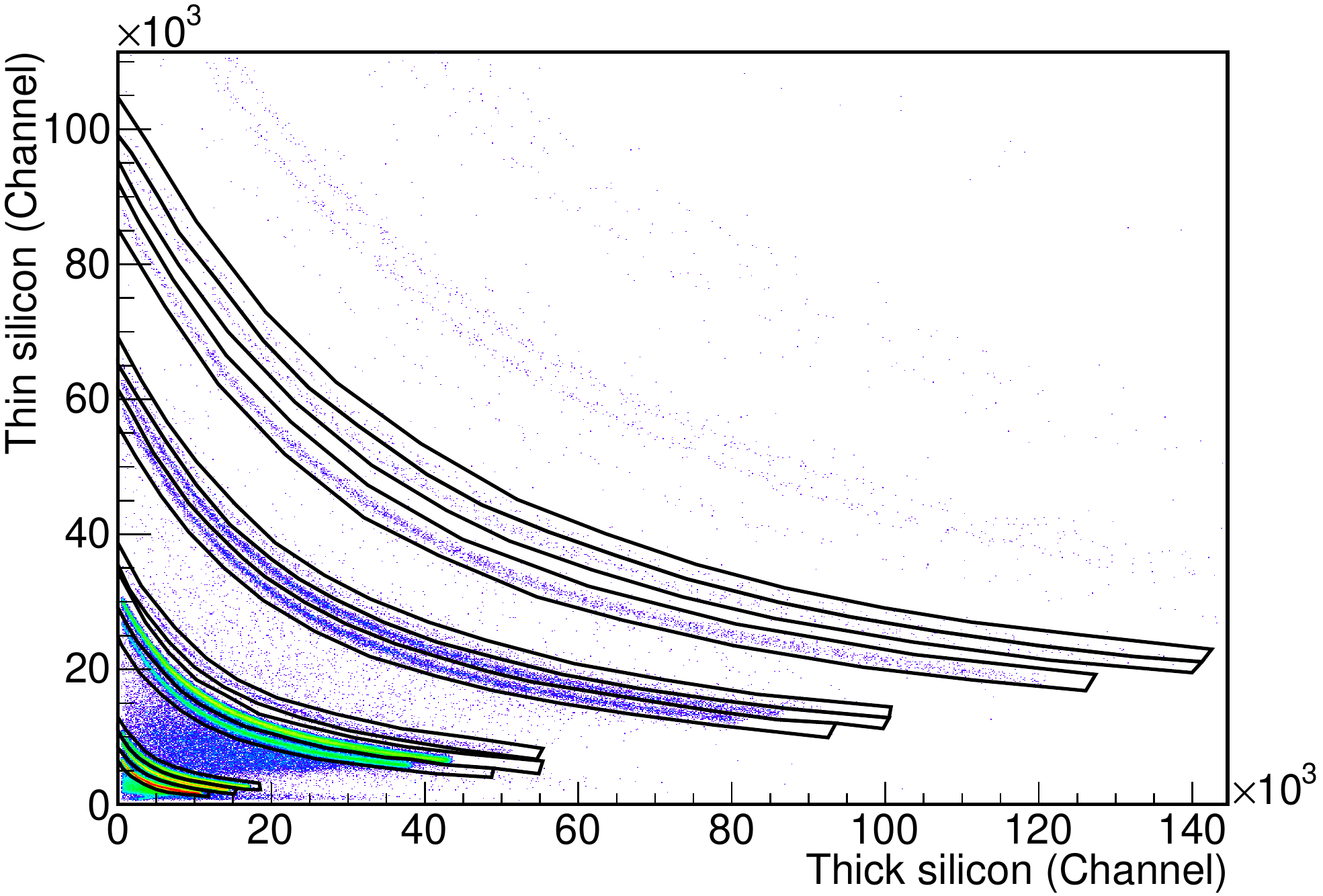}}
\caption{Identification map between the first stage and the second stage of a telescope. The black lines represent the different graphical contours used for the particles identification.}
\label{carteFinEpais}
\end{figure}

\begin{figure}[H]
\centerline{\includegraphics[width=0.7\textwidth]{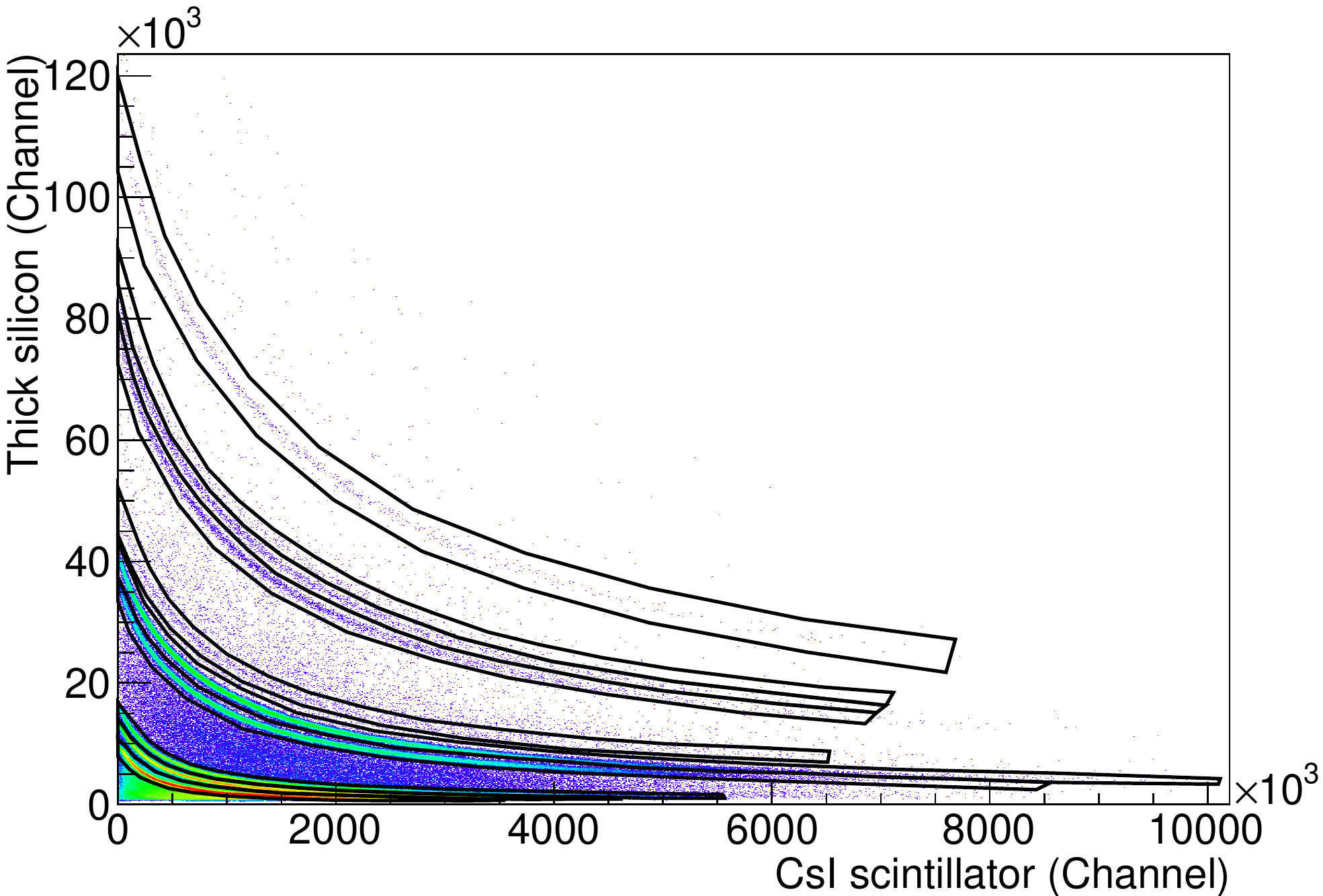}}
\caption{Identification map between the second stage and the third stage of a telescope.}
\label{carteEpaisCsI}
\end{figure}

The $\Delta$E-E technique is used to perform the identification of the charged particles. This method is based on the fact that the energy loss of charged particles in matter depends on their charge and mass. Figs.~\ref{carteFinEpais} and \ref{carteEpaisCsI} show examples of identification maps for the 250~$\mu$m thin Carbon target. Each hyperbole corresponds to a particle characterized by its charge and mass numbers (Z,A). Figure \ref{carteFinEpais} represents the energy deposited in the thin silicon detector relative to the energy deposited in the thick one. The hyperbolas back bending are due to the particles which cross the second stage of the telescope. Figure \ref{carteEpaisCsI} shows the energy deposited in the thick Si detector relative to the energy deposited in the CsI scintillator. This map is complementary to the previous one, thanks to the identification of the particles passing through the first two stages. The identification is performed by a set of contours for each telescope in order to separate each particle in charge and mass, from Z = 1 to Z = 6. These contours are represented by black lines on the maps of the figures \ref{carteFinEpais} and \ref{carteEpaisCsI}. Since the data acquisition is triggered when the second stage is hit, the identification threshold at low energy is determined by the thickness of the first stage. The minimum energy requisite to cross the first stage (150~$\mu$m) and thus to hit the second one depends on the nuclear charge of the particle, these threshold values are given in Table \ref{threshold}. The CsI crystal thickness prevents most of the charged particles from passing through the scintillator. Indeed, the lightest particles namely protons, deuterons and tritons, need respectively energies greater than 198, 130 and 103~MeV/u to cross the CsI crystal. 

\begin{table}[H]
\centering
\begin{tabular}{c c}
\hline\hline
Nuclear charge & Threshold (MeV)\\ 
\hline
1 & 4 \\
2 & 14 \\
3 & 30 \\
4 & 46 \\
5 & 61 \\
6 & 81 \\
\hline \hline
\end{tabular} 
\caption{Minimum energy for detection as a function of the charge of the particles.}
\label{threshold}
\end{table}

\subsection{Energy calibration} 
\label{calibration_didier}
\begin{figure}[H]
\centerline{\includegraphics[width=0.7\textwidth]{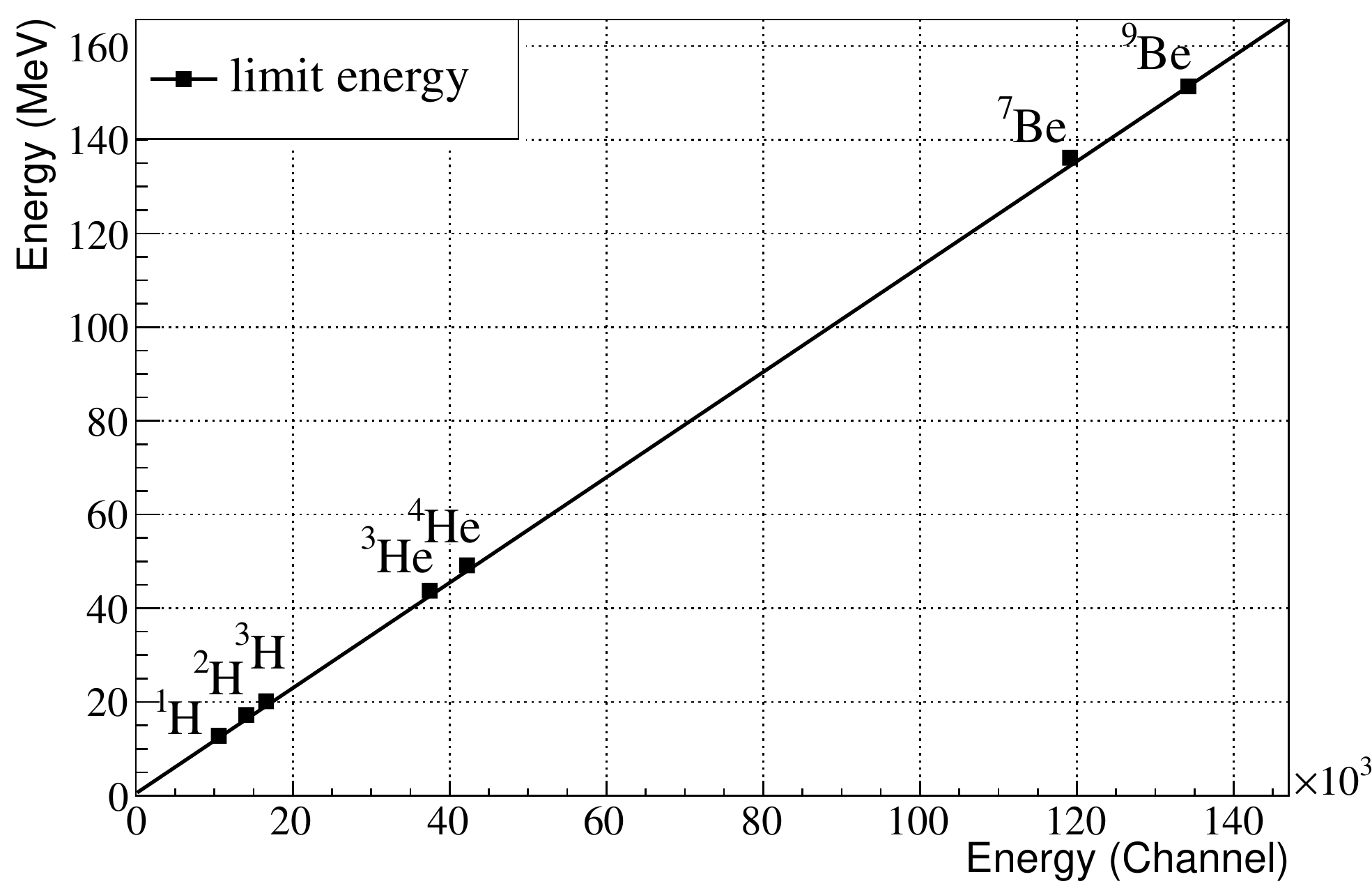}}
\caption{Energy calibration for the thick silicon detector of a telescope (linear fit).}
\label{etalSi}
\end{figure}

The energy calibration of the two silicon detectors is based on the maximal deposited energy in the second stage for a given particle without crossing it. Thanks to the known thicknesses of the Si detectors, this energy threshold is computed, for each particle, by using SRIM~\cite{srim} simulations and is linked to the corresponding channel number of the thick Si detector. The energy loss in the thin Si detector is deduced from the energy deposited in the thick Si detector. This allows to establish the correspondence between the energy deposited in the thin Si detector and the channel number in the corresponding charge to digital converter (QDC) module. Fig.~\ref{etalSi} shows an example of the calibration points obtained for one thick Si owing to this method. As the response of a silicon detector does not depend on the nature of the particle, the correlation between the energy loss and the channel number is the same for all particles. The calibration parameters are obtained by a linear fit of the points as shown on Fig.~\ref{etalSi}.

Contrary to silicon detectors, the produced light in the scintillator depends on the nature of the particle and on its energy. It is necessary to use a calibration function with a distinct set of parameters for each isotope. The chosen function, based on the one proposed by Stracener et al~\cite{Stracener 90}, is described by the following formula:

 \begin{equation}
\text{E}_{\text{MeV}} = \alpha + \beta \text{E}_{\text{channel}} + \gamma ln(1 + \delta \text{E}_{\text{channel}}) + \eta \text{E}^{2}_{\text{channel}}
\label{stracener}
\end{equation}

\begin{figure}[H]
\centerline{\includegraphics[width=0.7\textwidth]{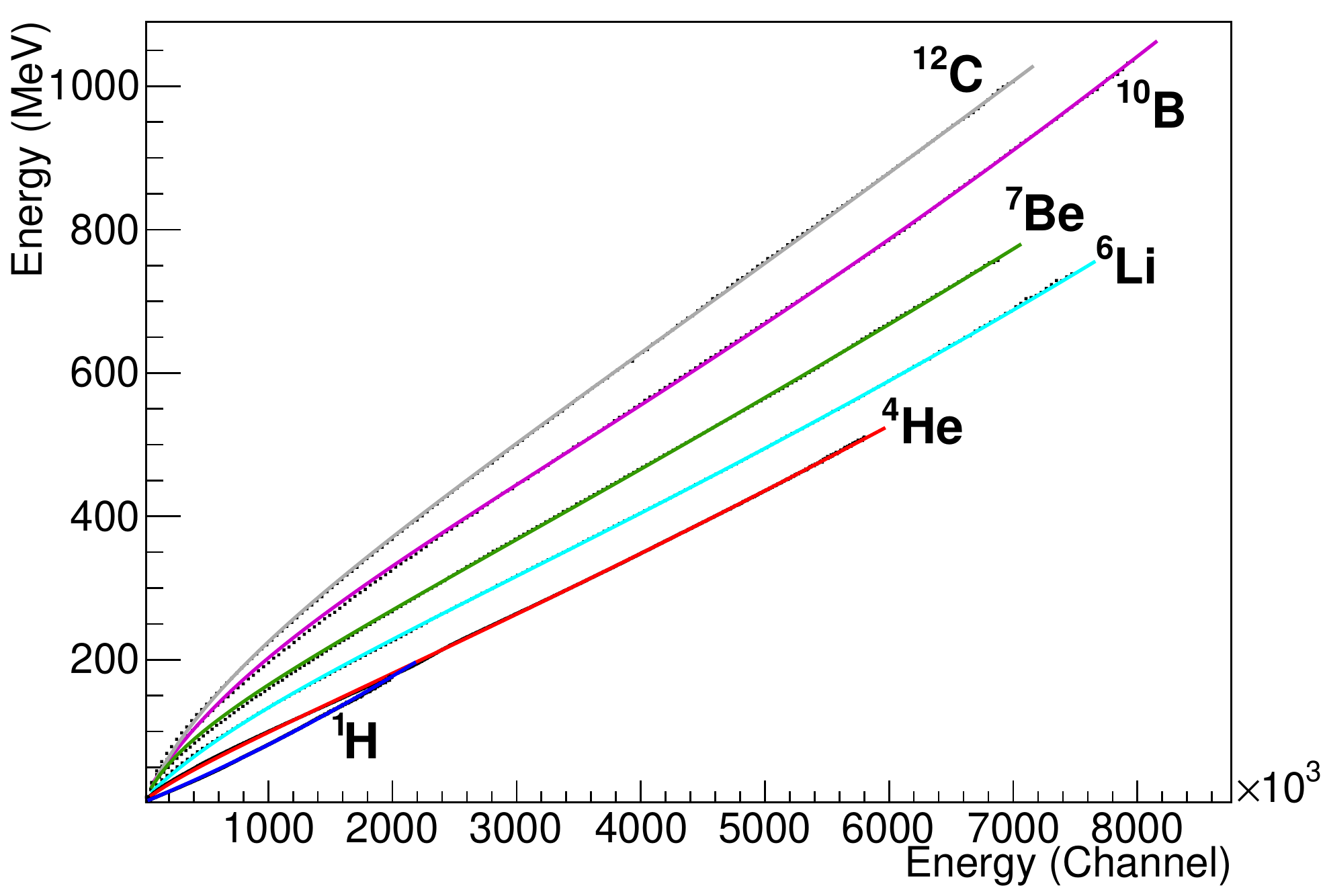}}
\caption{Calibration of the CsI scintillator of a telescope thanks to the calibration function given by Eq. \ref{stracener}.}
\label{etalCsI}
\end{figure}

This equation  takes into account the quenching of the scintillator at small energy as well as the linear response at higher energy. The term, $\eta$E$^{2}_{\text{channel}}$, has been added to take into account a possible saturation at very high energy. The energy loss in the CsI detector is deduced from the energy deposited in the thick Si detector using SRIM simulations. This method propagates the errors made on the energy calibration of the thick Si detector to the energy calibration of the CsI detector. The lower the deposited energy in the thick Silicon detector is, the larger the error on the residual energy released in the CsI is. With our experimental setup, this error has been estimated, from almost 30\% for 95~MeV protons to 4\% for 95~MeV/u carbon ions. An example of the calibration functions obtained is shown on Fig.~\ref{etalCsI}. The five parameters ($\alpha$, $\beta$, $\gamma$, $\delta$ and $\eta$) are determined for each isotope and for each telescope.

\section{Data Analysis using KaliVeda}

An new analysis method based on the KaliVeda toolkit \cite{kaliveda} has been used for the analysis of the experiment. KaliVeda is an object-oriented software, based on the ROOT~\cite{ROOT} framework for data processing, with many useful classes for data analysis and energy loss calculation. It has been developed to analyse data obtained by the $\Delta$E-E technique. It was firstly developed for the analysis of the INDRA multi-detector.

\subsection{KaliVeda tools}

In this part, the KaliVeda tools which are used for the analysis will be presented.

\subsubsection{The Energy Loss functional}

KaliVeda is using a functional describing accurately the energy loss $\Delta$E in a detector as a function of the residual energy E deposited in a second detector in which the particle has stopped. This functional is the one proposed by L.Tassan-Got in which the charges Z and the masses A of particles are parameters \cite{Tas02}. For detectors delivering a linear response (as silicon detectors), it reads:
\begin{equation}
\Delta \text{E} = [(g\text{E})^{\mu+\nu+1} + (\lambda \text{Z}^\alpha \text{A}^\beta)^{\mu+\nu+1} + \xi \text{Z}^2 \text{A}^\mu (g\text{E})^\nu]^{\frac{1}{\mu+\nu+1}} - g\text{E}
\label{tass_got}
\end{equation}
where $g,\mu,\nu,\lambda,\alpha,\beta,\xi$ are fit parameters.

For detector delivering a non linear response versus the deposited energy this functional needs to be corrected. For scintillators, it allows to take into account the quenching effect of scintillation leading to a non linearity of the emitted light versus the deposited energy. The energy E in equation~\ref{tass_got} must now be expressed as a function of the light $h$ emitted by the scintillator. The light response of CsI(Tl) crystals is deduced from the Birks formula \cite{Bir64} and allows to express the energy released in the CsI as a function of the emitted light :
\begin{equation}
\text{E} = \sqrt{h^2 + 2 \rho h \left[ 1 + ln \left(1 + \frac{h}{\rho}\right)\right] }
\label{tass-got-csi}
\end{equation}
where $\rho = \eta \text{Z}^2\text{A}$ is a new fit parameter.

\subsubsection{The KaliVeda identification grid}
\label{fit_grid_section}
As with the previous analysis method, the particle identification is performed using the $\Delta$E-E method. In such a representation, the particles are grouped around quasi-hyperbolas characterized by their charge and mass. Then, for each detected particle, the closest line has to be identified and the corresponding charge and mass have to be assigned to the particle. It can be done by using graphical contours as presented in the previous section or by using the KaliVeda identification tools.

\begin{figure}[H]
\centerline{\includegraphics[width=0.7\textwidth]{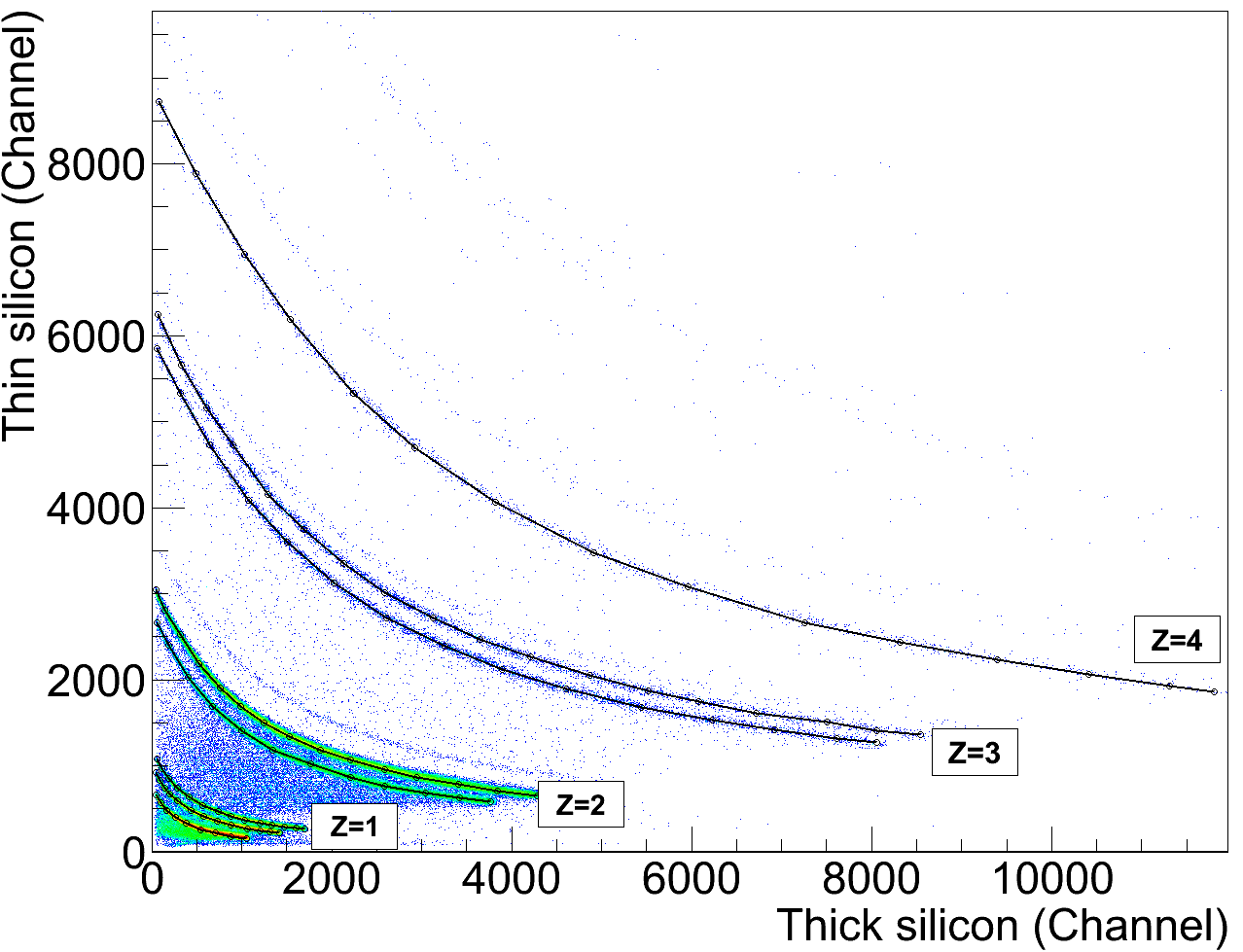}}
\caption{Handmade identification grid built on a $\Delta$E-E map for the most produced isotopes.}
\label{KVIDGrid}
\end{figure}

\begin{figure}[H]
\centerline{\includegraphics[width=0.7\textwidth]{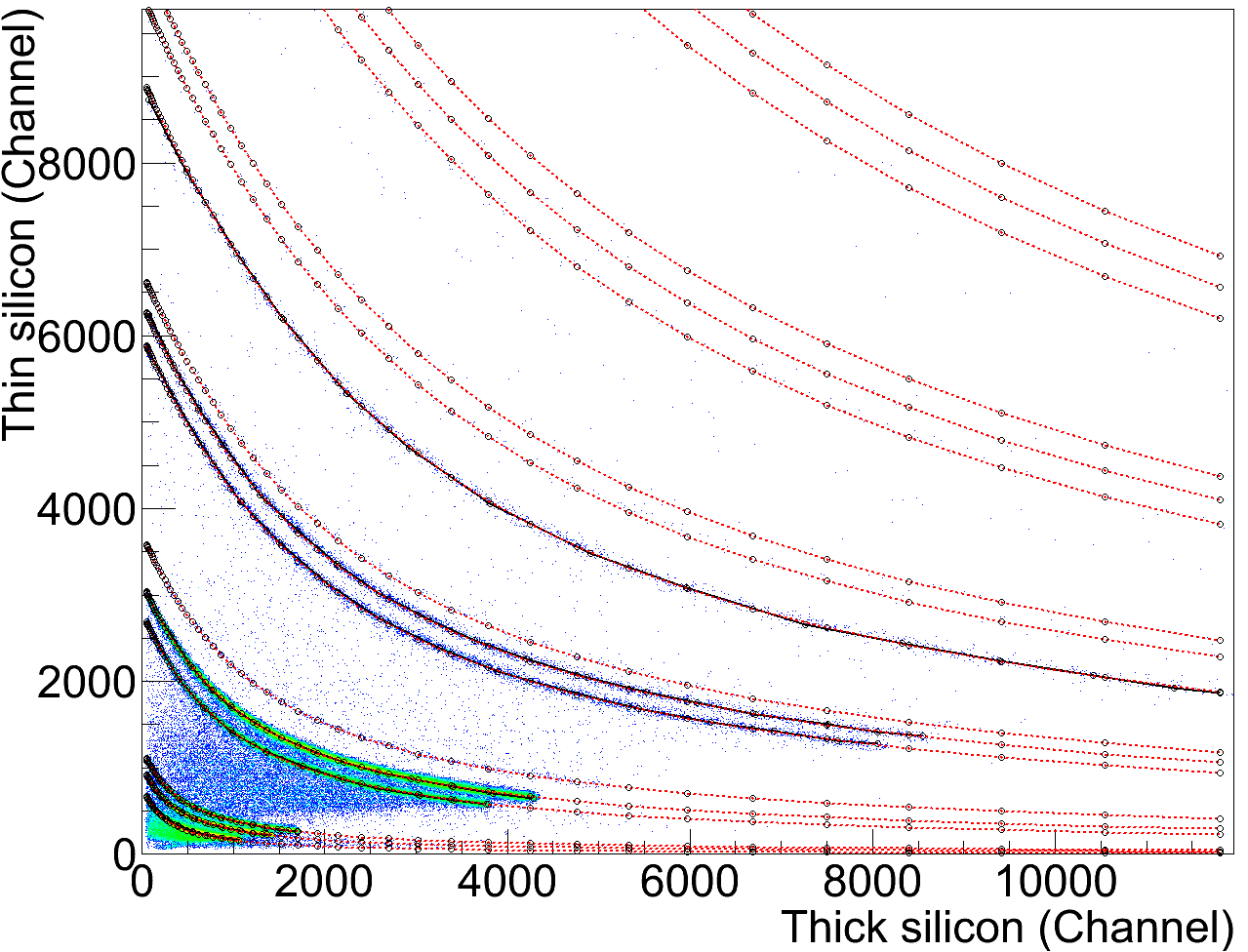}}
\caption{New identification grid obtained by fitting the Tassan-Got functional to the handmade identification grid.}
\label{fit_grid}
\end{figure}

KaliVeda is using identification grids which are graphs that contain a set of identification lines as described by equation~\ref{tass_got} (cf Fig.~\ref{KVIDGrid} and Fig.~\ref{fit_grid}). Each line corresponds to an isotope and reproduces the shape of its quasi-hyperbola on the $\Delta$E-E map. 

In a first step, to construct the grids, the user has to do a ``handmade identification grid'' for a few lines on the $\Delta$E-E map. To do so, one has to give the coordinates of about ten points of the relevant lines and to specify charge and mass of the corresponding particle as shown on Fig. \ref{KVIDGrid}. Only few lines are needed here (those with the highest statistics), whereas the other one will be computed afterwards. 

In a setup with three stages telescopes, two identification grids are needed per telescope (one for the particles stopped in the thick Silicon detector and one for those stopped in the CsI). 
Fig.~\ref{KVIDGrid} represents the "handmade identification grid'' for the particles stopped in the thick Silicon detector (superimposed to the $\Delta$E$_\text{thin}$-E$_\text{thick}$ map).

In a second step, the functional of equation~\ref{tass_got} is fitted to the handmade $\Delta$E$_\text{thin}$-E$_\text{thick}$ identification grids. The same method is used for the $\Delta$E$_\text{thick}$-E$_\text{CsI}$ identification grids but the energy deposited in the CsI is here expressed as a function of the emitted light as described in equation~\ref{tass-got-csi}. The fit result allows to obtain the analytical expression of the identification lines for all isotopes. New identification grids including the lines of isotopes with low statistics can now be created, as shown on Fig~\ref{fit_grid}.

By this identification method, two sets of accurate $\Delta$E-E identification grids ($\Delta$E$_\text{thin}$-E$_\text{thick}$ and $\Delta$E$_\text{thick}$-E$_\text{CsI}$) have been determined. 

\subsection{The KaliVeda method}

In this part, the method we have developed will be presented. This method is using the KaliVeda tools described above.

\subsubsection{Particle identification}

\begin{figure}[H]
\centerline{\includegraphics[width=0.7\textwidth]{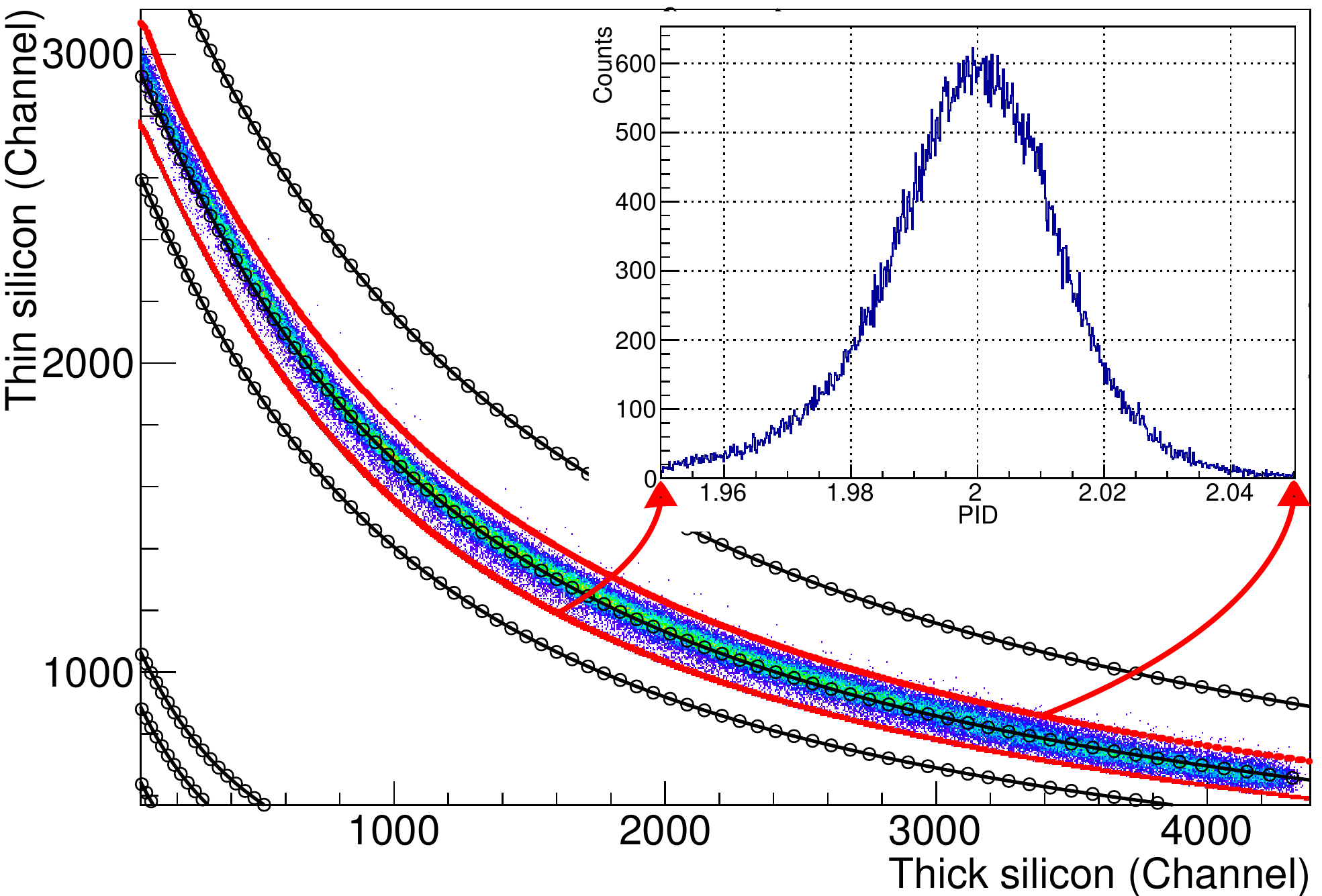}}
\caption{Example of PID calculation for $^4$He fragments.}
\label{PID1}
\end{figure}

\begin{figure}[H]
\centerline{\includegraphics[width=0.7\textwidth]{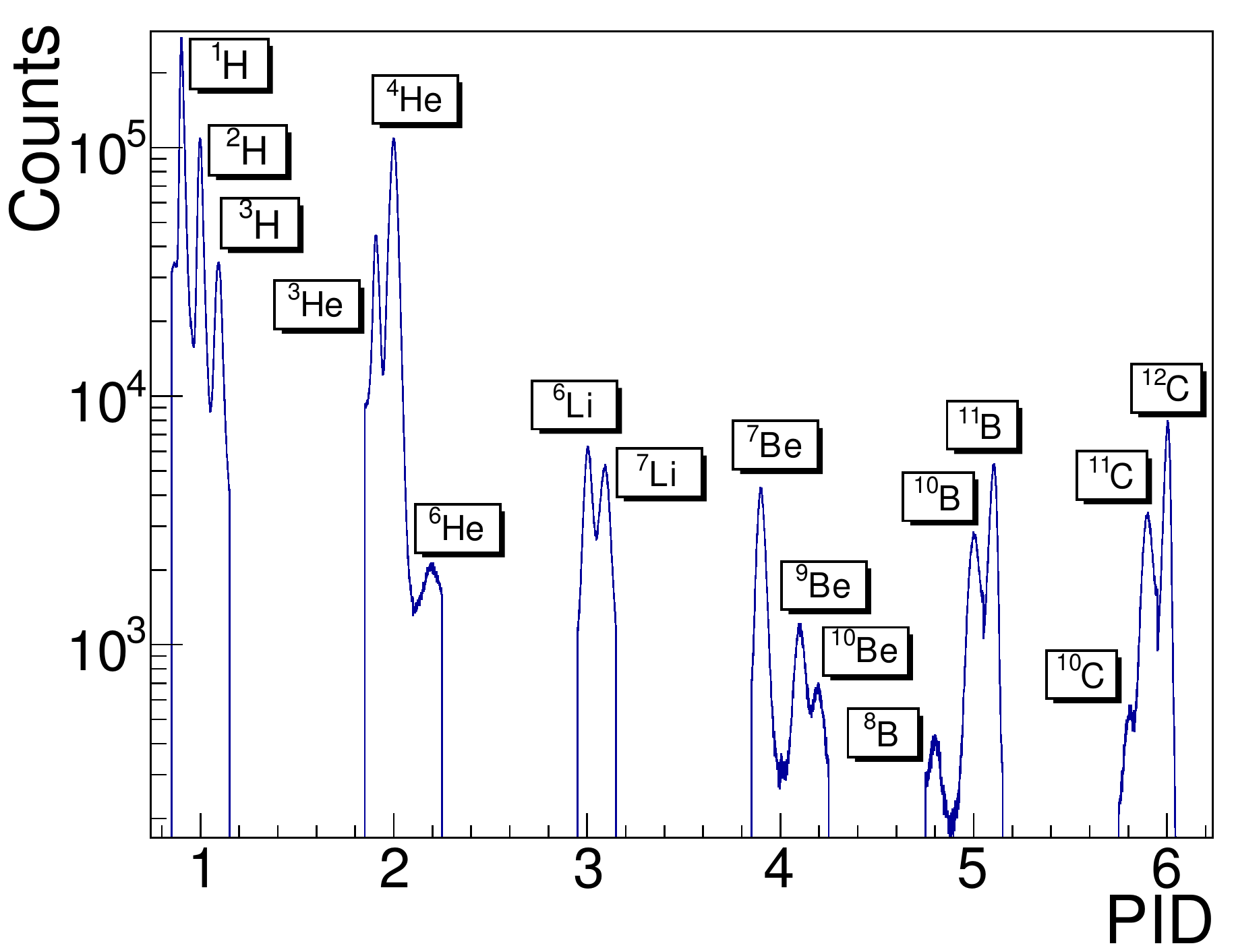}}
\caption{PID calculation for all detected isotopes.}
\label{PID2}
\end{figure}

To identify the detected particles, the procedure works as follows: 
\begin{itemize} 
 \item For each experimental measurement, the associated identification grid is loaded and the algorithm computes the distance on the $\Delta$E-E map between the coordinates of the detected particle (($\Delta$E$_\text{thin}$;E$_\text{thick}$) or ($\Delta$E$_\text{thick}$;E$_\text{CsI}$)) and the different isotope's lines. 
 \item The charge and mass assigned to the particle are the one corresponding to the closest line of the identification grid. 
 \item A new variable named A$_{real}$ is determined as a function of the distance to the isotope's line with A$_{real} \in [$A$-0.5;$A$+0.5]$. The value of A$_{real}$ is close to the mass number A when the experimental point is close to the identification line. The minimal or maximal values of A$_{real}$ correspond to points located at the same distance of two lines with consecutive mass numbers.
 \item This new variable is used to determine the Particle Identifier (PID) expressed as PID $ = $ Z + 0.1 $\times($A$_{real}-2\times$Z$)$. 
\end{itemize}
 As examples, the PID for $\alpha$ particles may range from 1.95 to 2.05, the PID of $^3$He isotopes may ranges from 1.85 to 1.95 and the PID for $^6$He may range from 2.15 to 2.25.
 Fig.~\ref{PID1} illustrates an example of PID calculation for alpha particles. The result of the PID calculation for all the detected isotopes is shown on Fig.~\ref{PID2}. The quality of the identification is very good and the different isotopes are clearly separated, even those with low statistics.
 
\subsubsection{Energy calibration}

Once the particle identification has been achieved, the energy calibration of the detectors can be performed.

For energy calibration of the Silicon detectors, the analytical expressions~(eq.~\ref{tass_got}) of the identification grids for the $\Delta$E$_\text{thin}$-E$_\text{thick}$ map are used.
First, a theoretical identification grid is generated. The points of this grid are obtained with the energy loss calculation tool of KaliVeda, taking into account the Silicon detectors thicknesses. The energy loss functional is then fitted to the theoretical identification grid. It gives an analytical expression of the theoretical lines. Then, the analytical expression of the experimental lines are obtained as described in part~\ref{fit_grid_section}.

\begin{figure}[H]
\centering
\subfigure[]{{\includegraphics[width=0.49\textwidth]{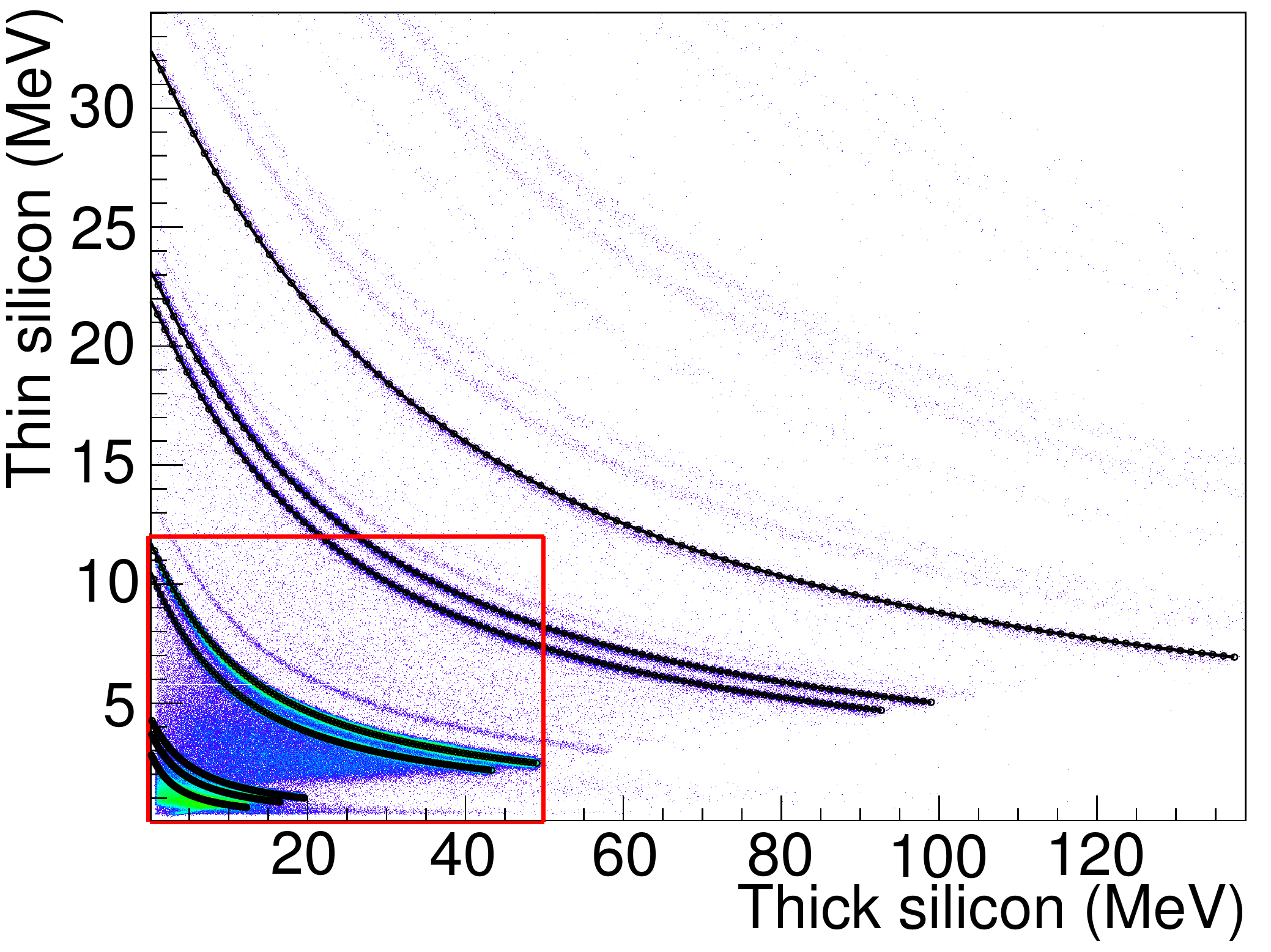}}}
\subfigure[]{{\includegraphics[width=0.49\textwidth]{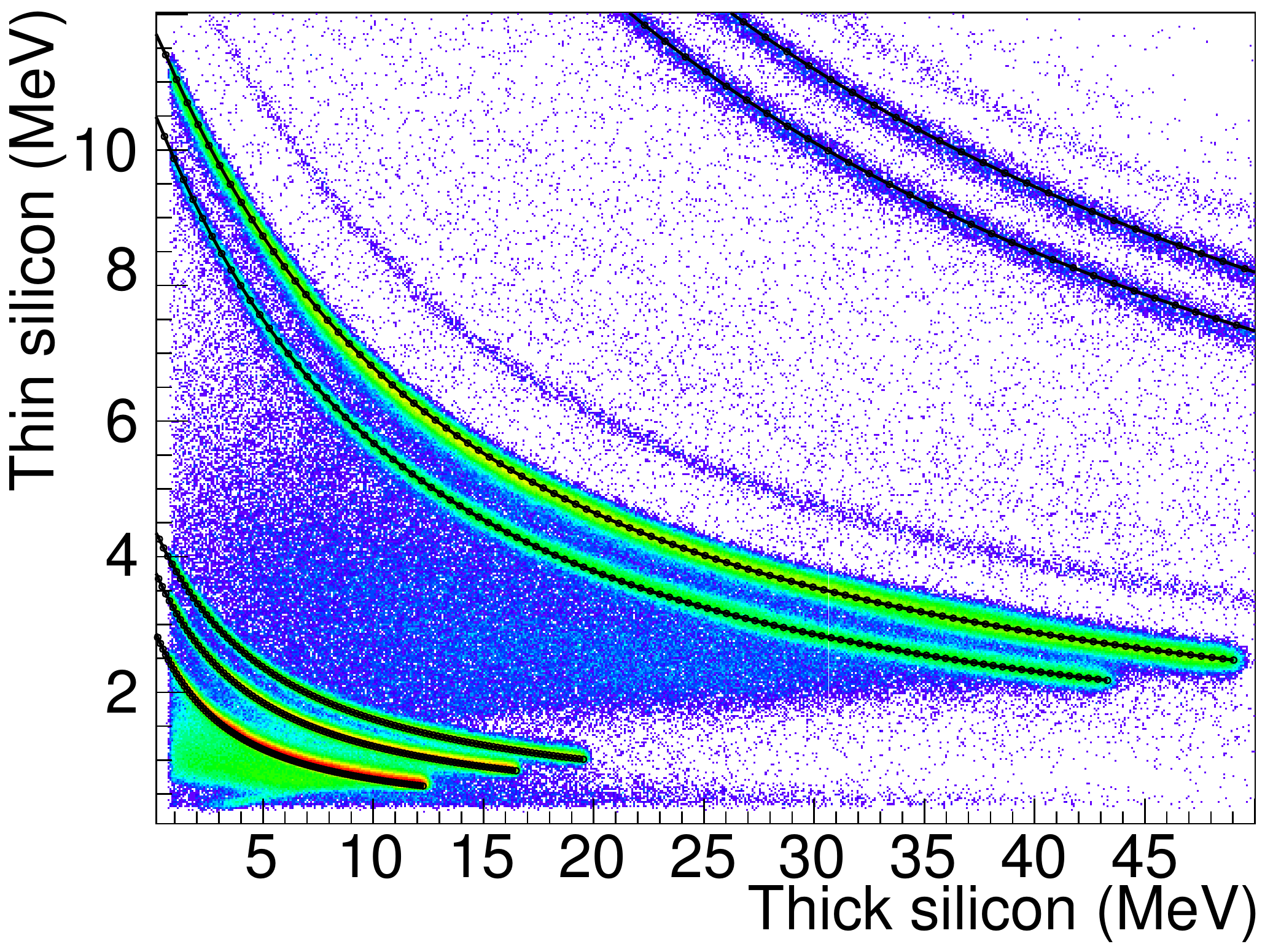}}}
\caption{Result of the Silicon detectors calibration. The theoretical grid is superimposed to the $\Delta$E-E calibrated map (the right-hand side is a zoom of the red square). Only statistical errors are represented.}
\label{calib}
\end{figure}

Two sets of parameters have thus been obtained, one for the theoretical grid expressed in MeV, and one for the experimental one expressed in channels. As Silicon detectors are delivering a linear response, their energy calibration consists in determining the affine relation between channels and MeV. Four calibration parameters have to be determined, two for the thin Silicon detector and two for the thick Silicon detector. By applying the correct transformations respectively to $\Delta$E and E in the experimental energy loss functional, the two functionals (experimental and theoretical) will be equal. This procedure is done by using an iterative algorithm which adjusts the four calibration parameters to make the two functionals as close as possible. Fig.~\ref{calib} represents a $\Delta$E-E calibrated map on which the theoretical grid is superimposed. Both are in perfect agreement.

The CsI energy calibration can not be done with the same method because of the non linear response of the emitted light versus the deposited energy (cf eq. \ref{tass-got-csi}). The thick Silicon detector energy calibration is used to determine the deposited energy in the CsI as follows. Once a particle is identified in the $\Delta$E$_\text{thick}$-E$_\text{CsI}$ map, its charge, mass and energy loss in the thick Silicon detector are known. Then, a KaliVeda tool computes the corresponding incident energy of the particle thanks to the known energy loss in the Silicon detector. The residual energy which will be deposited in the CsI can thus be deduced. The disadvantage of such a method, as the one used in section \ref{calibration_didier}, is that the error made on the thick Silicon detector calibration parameters is propagated to the energy deposited in the CsI.  This error has been estimated, in this case, from around 20\% for 95~MeV protons to 4\% for 95~MeV/u $^{12}$C.

\medskip

The main advantage of using the KaliVeda toolkit is that once the identification grids are created, they allow the detector calibration and the particle identification at the same time. With such a method, the energy calibration is done not only on a few points but for all the points of the identification lines. Moreover, these procedures are automated which saves time in data analysis. The time required, for the user, to construct the necessary elements for the calibration and identification is about one hour per telescope.

\section{Comparison of the two analysis methods}

The obtained data have been analyzed with the two methods described above. In this part, some results of fragmentation cross section measurements for the carbon target will be presented and the two analysis methods will be compared.

Firstly, the angular distributions, obtained for angles from $4^\circ$ to $43^\circ$, will be compared. The $^{12}$C fragmentation cross sections for a $^\text{A}_\text{Z}$X fragment are obtained as follows :

\begin{equation}
\frac{d\sigma}{d \Omega} (^\text{A}_\text{Z}\text{X}) = \frac{\text{N}_{^\text{A}_\text{Z}\text{X}} \times \text{A}_{\text{target}}}{\text{N}_{^{12}\text{C}} \times \Omega \times \rho \times th \times \mathcal{N}_{\text{A}}}
\label{cross_section}
\end{equation}

where N$_{^\text{A}_\text{Z}\text{X}}$ is the number of $^\text{A}_\text{Z}\text{X}$ fragments detected, $\text{A}_{\text{target}}$ is the target mass, N$_{^{12}\text{C}}$ is the number of incident carbon nuclei, $\Omega$ is the solid angle of the detector, $\rho$ and $th$ are the target density and thickness respectively and $\mathcal{N}_{\text{A}}$ is the Avogadro's number.

\begin{figure}[H]
\centerline{\includegraphics[width=0.7\textwidth]{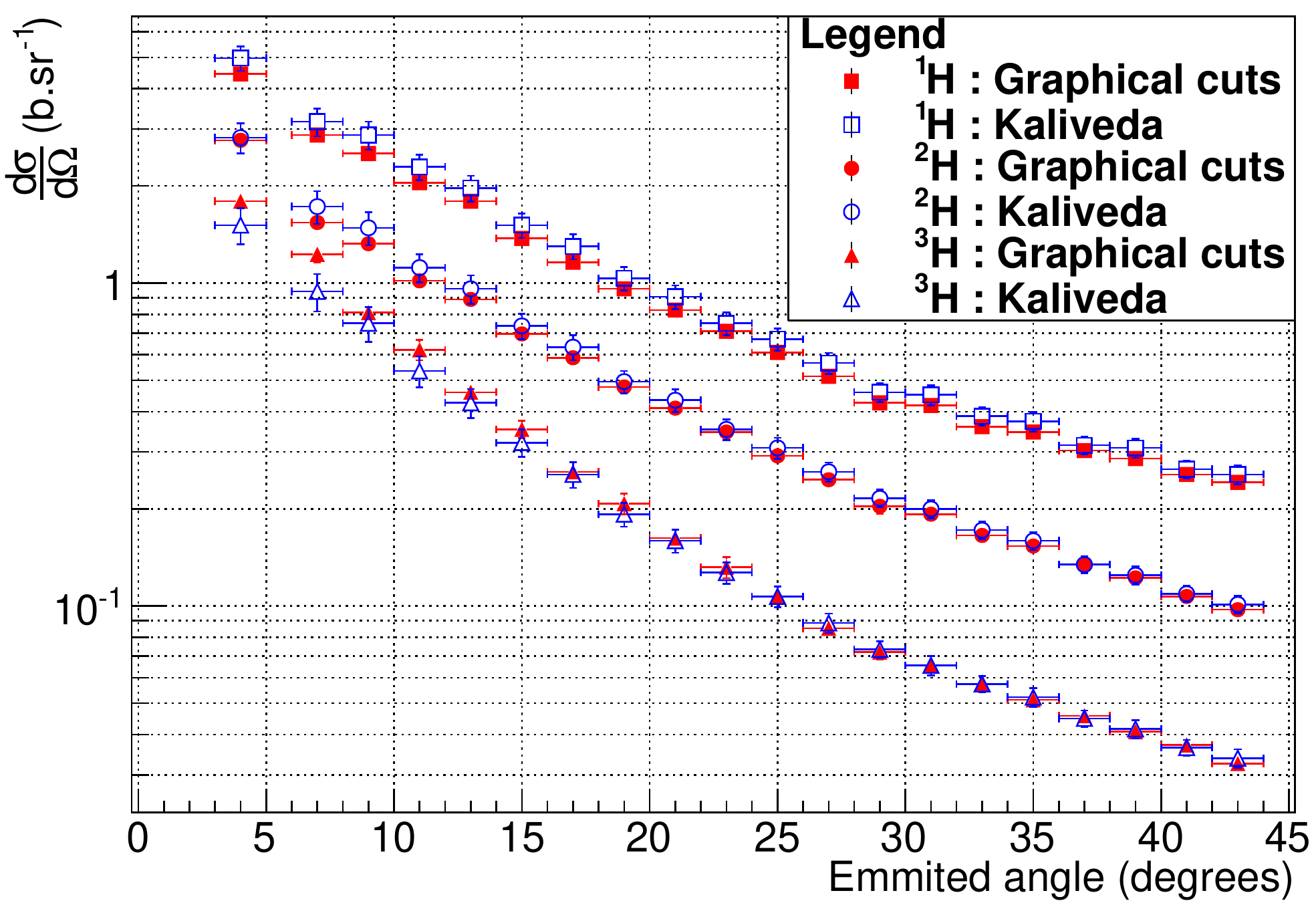}}
\caption{Comparisons of the Hydrogen angular distributions obtained with the two analysis methods. Only statistical errors are represented.}
\label{theta_protons}
\end{figure}

\begin{figure}[H]
\centerline{\includegraphics[width=0.7\textwidth]{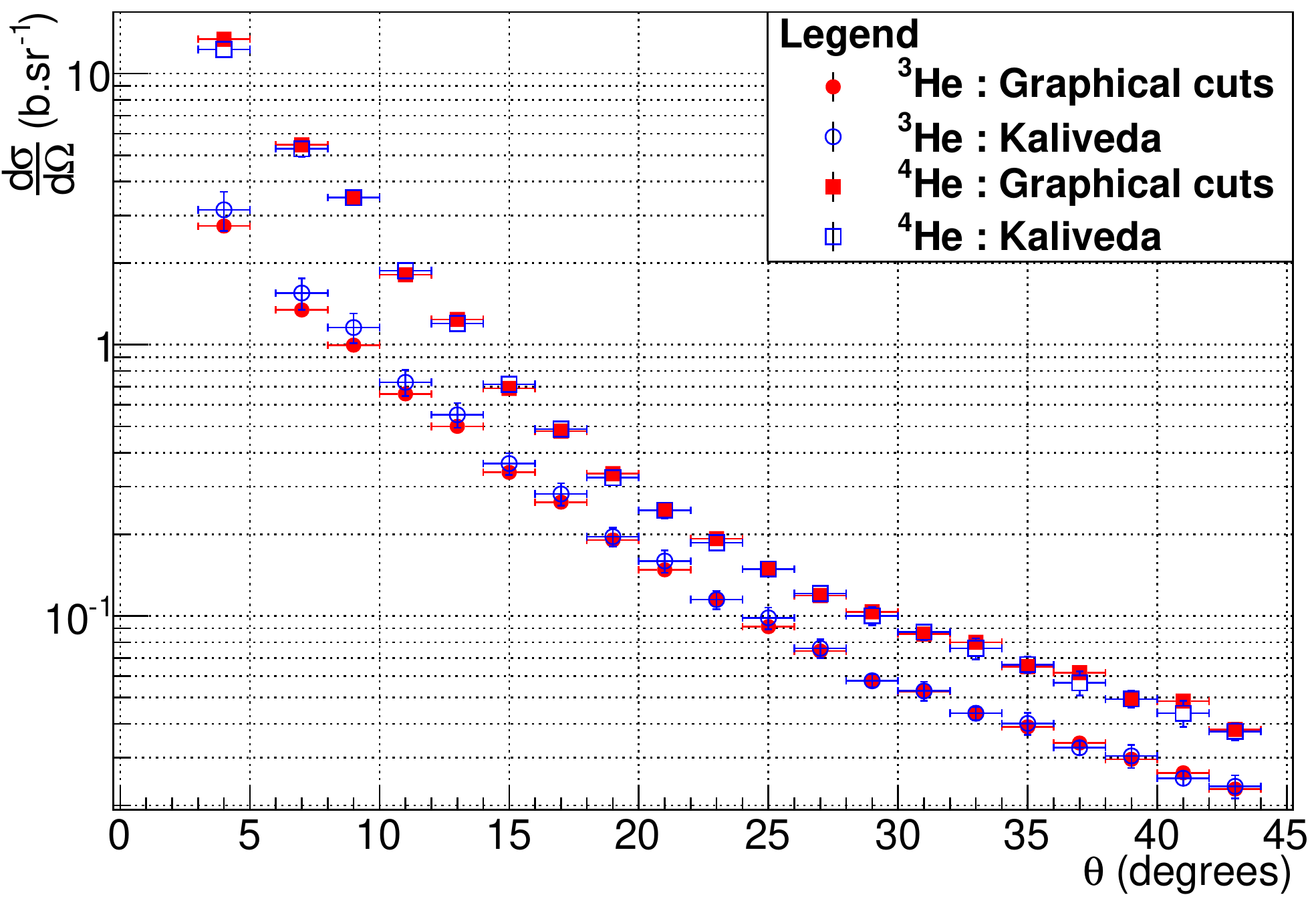}}
\caption{Comparisons of the Helium angular distributions obtained with the two analysis methods. Only statistical errors are represented.}
\label{theta_alphas}
\end{figure}

\begin{figure}[H]
\centerline{\includegraphics[width=0.7\textwidth]{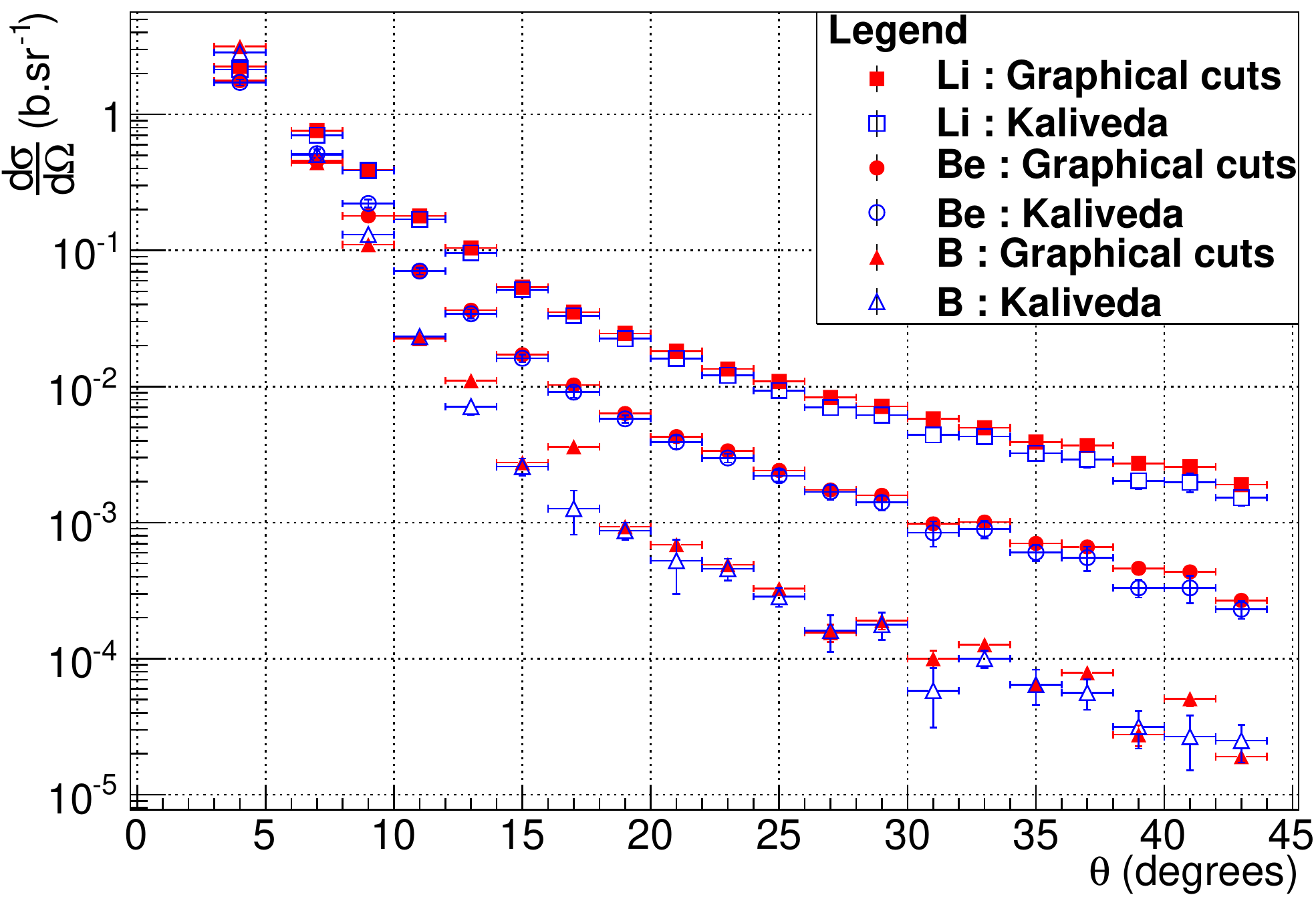}}
\caption{Comparisons of the angular distributions obtained with the two analysis methods for the heavier fragments. The Z distributions are obtained by adding the cross sections of each isotope.}
\label{theta_lourds}
\end{figure}

Fig.~\ref{theta_protons}, Fig.~\ref{theta_alphas} and Fig.~\ref{theta_lourds} display the angular distributions obtained with the two analysis methods for different emitted fragments. Statistical errors only are represented, systematic errors are still under study. As expected, the fragment emission is dominated by H and He isotopes with a predominance of $\alpha$  at low angles (below $\sim 10^\circ$) which is compatible with the alpha cluster structure of the $^{12}$C. The results also show an angular emission most forward peaked for heavier fragments. 

The two analysis methods give consistent results. The discrepancies between the two methods are of $\pm~5\%$ for most of the isotopes and angles. However, the discrepancies can reach 15\% at low angles for light isotopes or for isotopes with low statistics. These differences are due to the widths of the particle selection cuts which are not the same for the two methods (width of the contours for the first method and width of the identification lines for the second one). The differences are more important for protons because of the increasing "background" on the $\Delta$E-E maps at low energy. As this "background" is mainly due to protons escaping the CsI by scattering, a large identification line width has been used in the KaliVeda method in order to not lose proton statistics. 

\medskip

\begin{figure}[H]
\centering
\subfigure[Proton energy distributions]{\label{E_proton}{\includegraphics[width=0.49\textwidth]{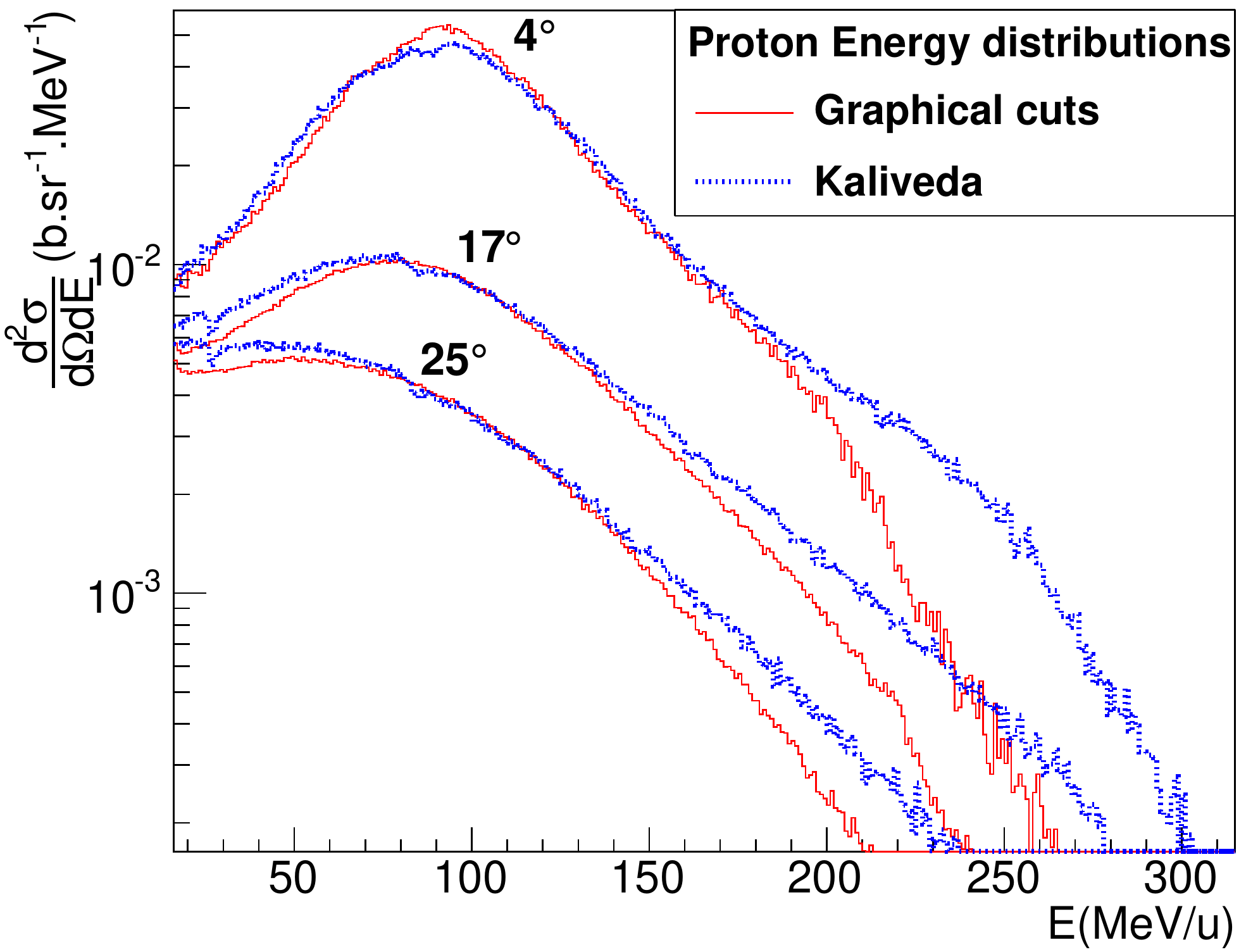}}}
\subfigure[$\alpha$ energy distributions]{\label{E_alpha}{\includegraphics[width=0.49\textwidth]{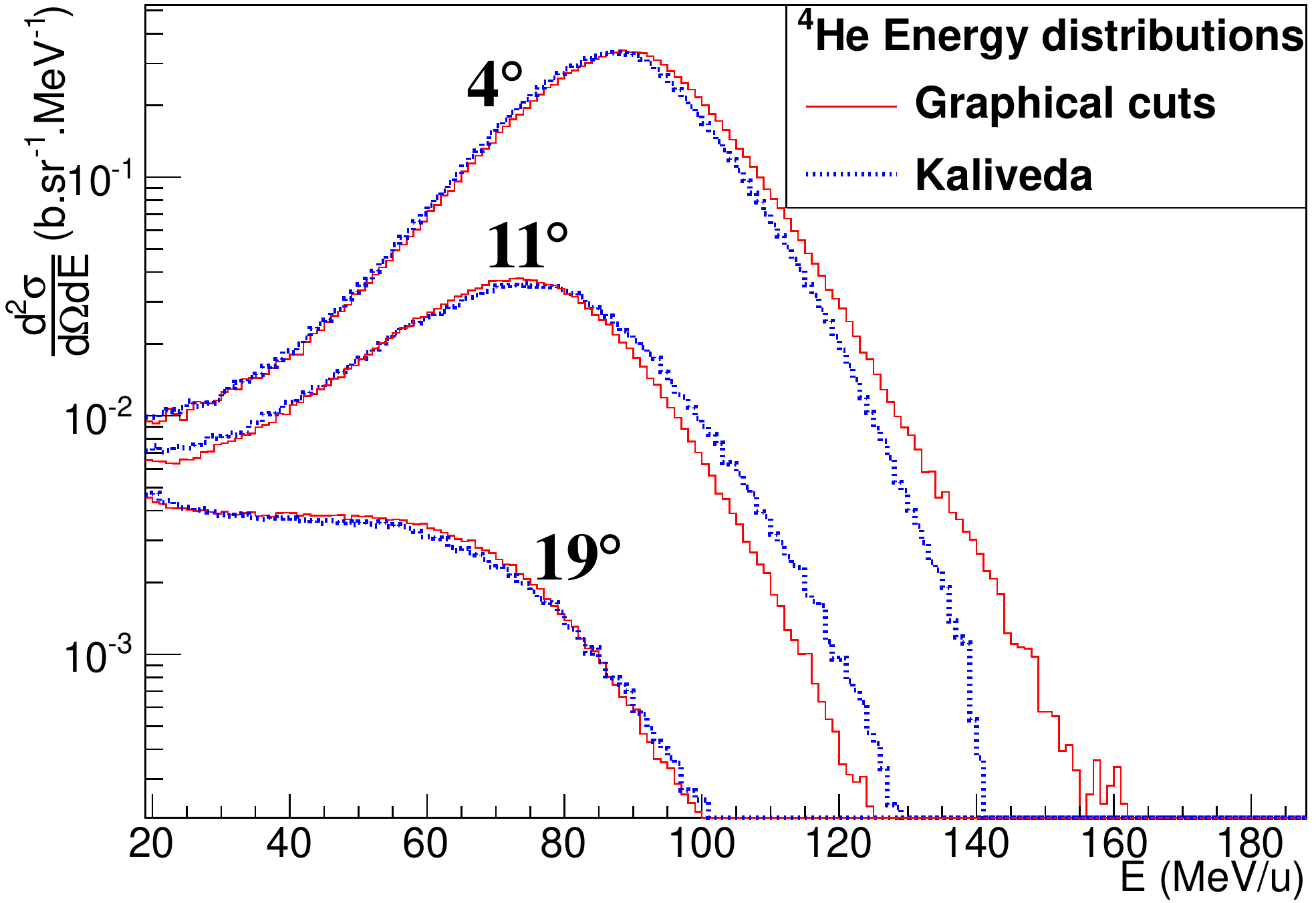}}}
\caption{Comparisons of the energy distributions obtained with the two analysis methods for protons and $\alpha$ at different angles.}
\label{energy_dist}
\end{figure}

Secondly, the energy distributions will be compared. Fig.~\ref{energy_dist} represents the energy distributions of protons and $\alpha$ fragments at different angles obtained with the two analysis methods. These distributions are dominated at low angles by a peak centered close to the beam energy (95 MeV/u). The energy and amplitude of this peak are decreasing with the emission angle. This observation confirms that most of the detected fragments are coming from the projectile fragmentation.

The energy distributions are also in good agreement. The discrepancies between the two methods are due on one hand to the propagation of the energy calibration of the thick Silicon detector. This effect can be seen on Fig. \ref{E_alpha} in which the differences between the two analysis method are increasing with the energy. It should be noticed that the discrepancies stay under 10\% even for high energy protons. On the other hand, the discrepancies are also coming from the particle selection width as explained for the angular distribution. This effect is clearly visible for the higher energies of Fig. \ref{E_proton} in which the largest width of the KaliVeda methods leads to the selection of background (due to the presence of $^2$H and $^3$H). As the background of the experiment is low, this effect is statistically low even with the choice of a selection width larger in the KaliVeda method than in the classical one. The comparisons between the energy distributions obtained with the two analysis methods show the coherence and the robustness of the two identification and calibration methods.

\section{Conclusion}

Two different methods for the data analysis of impinging $^{12}$C ions at 95~MeV/u on thin carbon target have been presented in this paper. Measurements for twenty different angles led to production rates, angular distributions and energy spectra for fragmented particles, ranging from protons to carbon ions. Two different analysis methods have been used, the first one is based on graphical cuts while the second one uses the KaliVeda framework. Both of them are in good agreement, within a discrepancy close to 5\% for most angular distributions, the discrepancy is higher concerning energy distributions in some specific cases. Angular distributions are dominated by the emission of light fragments (Z $<$ 3), more especially by $\alpha$  particles at forward angles, that is consistent with the $\alpha$ cluster structure of the $^{12}$C ion. Energy distributions dominated at low angles by a peak close to the beam energy indicate that most of the emitted particles result from the projectile fragmentation during nuclear reactions. As mentioned above, other targets have been used in this experiment using the same beam at the same energy, which will be the subject of an upcoming paper. The analysis of these data will help to determine the double differential cross section for the different couples C-H, C-O, C-Al and C-Ti.


\end{document}